\documentclass[letterpaper]{article} 

\usepackage{aaai2026}
\nocopyright

\usepackage{times}  
\usepackage{helvet}  
\usepackage{courier}  
\usepackage[hyphens]{url}  
\usepackage{graphicx} 
\usepackage{natbib}  
\usepackage{caption} 
\usepackage{algorithm}
\usepackage{algorithmic}
\usepackage{amsmath}
\usepackage{amssymb}
\usepackage{booktabs}
\usepackage{array}
\usepackage{makecell} 
\usepackage{rotating}
\usepackage{multirow}
\usepackage{multicol}
\usepackage{threeparttable}
\usepackage{subcaption}
\usepackage{siunitx}
\usepackage{bibentry}

\usepackage{newfloat}
\usepackage{listings}
\DeclareCaptionStyle{ruled}{labelfont=normalfont,labelsep=colon,strut=off} 
\floatstyle{ruled}
\newfloat{listing}{tb}{lst}{}
\floatname{listing}{Listing}
\def\showauthors@on{T}

\title{Distillation-Enhanced Clustering Acceleration for Encrypted Traffic Classification}
\author {
    Ziyue Huang\textsuperscript{\rm 1 2}\footnotemark[1],
    Chungang Lin\textsuperscript{\rm 1 2}\footnotemark[1],
    Weiyao Zhang\textsuperscript{\rm 1},
    Xuying Meng\textsuperscript{\rm 1}\footnotemark[2],
    Yujun Zhang\textsuperscript{\rm 1 2}\footnotemark[2]
}
\affiliations {
    \textsuperscript{\rm 1}Institute of Computing Technology, Chinese Academy of Sciences\\
    \textsuperscript{\rm 2}University of Chinese Academy of Sciences    
}

\begin{document}

\maketitle

\footnotetext[1]{Both authors contributed equally to this research.}
\footnotetext[2]{Corresponding author.}

\begin{abstract}
Traffic classification plays a significant role in network service management. The advancement of deep learning has established pretrained models as a robust approach for this task. However, contemporary encrypted traffic classification systems face dual limitations. Firstly, pretrained models typically exhibit large-scale architectures, where their extensive parameterization results in slow inference speeds and high computational latency. Secondly, reliance on labeled data for fine-tuning restricts these models to predefined supervised classes, creating a bottleneck when novel traffic types emerge in the evolving Internet landscape. To address these challenges, we propose NetClus, a novel framework integrating pretrained models with distillation-enhanced clustering acceleration. During fine-tuning, NetClus first introduces a cluster-friendly loss to jointly reshape the latent space for both classification and clustering. With the fine-tuned model, it distills the model into a lightweight Feed-Forward Neural Network model to retain semantics. During inference, NetClus performs heuristic merge with near-linear runtime, and valid the cluster purity with newly proposed metrics ASI to identify emergent traffic types while expediting classification. Benchmarked against existing pretrained methods, NetClus achieves up to 6.2x acceleration while maintaining classification degradation below 1\%.
\end{abstract}

\section{Introduction}

Traffic classification, which categorizes flows by services type (e.g., video streaming, file transfer) or application providers (e.g., Youtube, Facebook, Netflix), is foundational to network management, enabling Quality of Service (QoS) provisioning, bandwidth optimization, and security enforcement. Research in encrypted traffic classification has aroused much attention and undergone substantial evolution.


Initial ML models, e.g., Flowprint \cite{flowprint} and AppScanner \cite{appscaner}, are robust to encryption; however, their accuracy is constrained by manual feature engineering, which leaves a high requirement on expert experience. Deep learning models \cite{Datanet} have revolutionized traffic classification by automating feature extraction from raw packet data, eliminating the manual feature engineering required in traditional machine learning approaches. Convolutional Neural Networks \cite{FlowPic} demonstrate significant advantages in capturing spatial patterns within packet payloads, enabling effective application identification through discriminative feature learning. Similarly, recurrent architectures like BiLSTM leverage sequential dependencies in traffic flows \cite{LSTM_Att}. These methods collectively outperform statistical approaches by learning complex traffic patterns directly from raw bytes. Nevertheless, DL models exhibit notable limitations, especially heavy dependence on large-scale labels, limiting their generalization for real-world scenarios. 


The evolution of pre-trained models (PTMs) from natural language processing to encrypted traffic classification represents a paradigm shift \cite{etbert, yatc, NetGPT, TraGe, trafficformer}. They leverage self-supervised pre-training on large-scale unlabeled data and supervised fine-tuning for downstream tasks with small-scale labels. 
ET-BERT \cite{etbert} adapts BERT's architecture for traffic bursts using masked language modeling and next sentence prediction to learn packet relationships; 
YaTC \cite{yatc} innovates with graph-based flow representations processed as images via masked image modeling; and TrafficFormer \cite{trafficformer} enhances parameter efficiency with BERT-aligned encoders and strategic data augmentation during fine-tuning.
These models advance encrypted traffic analysis by learning transferable representations from unlabeled traffic, while significantly eliminating the burdens for exhaustive labeling.

Despite substantial progress in encrypted traffic classification, PTMs confront two persistent limitations. First, the considerable scale of PTMs results in slow inference speeds and elevated latency under high-volume network traffic, impeding practical processing capabilities. Second, current PTMs of encrypted traffic still rely on supervised fine-tuning for classification, rendering them incapable of dynamically recognizing novel network protocols, cryptographic methods, or application categories in rapidly evolving internet ecosystems. These constraints highlight critical gaps in adaptability and inference efficiency for contemporary network environments.

These limitations are non-trivial to solve. To handle the slow inference problem, we observe that flows belonging to the same category tend to share similar features and naturally cluster together. This observation motivates collective inference, however, high-dimensional PTM features remain costly. Additionally, to handle unseen traffic categories, it often requires additional novelty detectors, continual retraining, or extensive human labeling, introducing substantial overhead and limiting classification efficiency. Therefore, a more efficient and lightweight solution is demanded.

To address these challenges, we propose NetClus, an efficient and lightweight framework that enhances existing pre-trained models by introducing {clustering-aware} knowledge {distillation} and hybrid inference acceleration. 
In \textit{clustering-aware knowledge distillation}, we first design a clustering-oriented loss for fine-tuning to reshape the PTM’s representation space, enhancing inter-class separability and improving the clustering quality of traffic features. To reduce inference latency, NetClus distills both semantic embeddings and prediction knowledge from the original PTM into a compact feed-forward network, enabling efficient classification on high-purity traffic clusters. Meanwhile, a \textit{hybrid inference acceleration} mechanism is introduced, i.e., flows from fine-grained high-purity clusters are directly mapped to true labels via pseudo-label matching, while ambiguous cases from few low-purity clusters are processed by the original PTM. This dual-path mechanism not only accelerates inference but also facilitates the discovery of novel or previously unseen traffic patterns through cluster-level analysis.

NetClus is designed to be model-agnostic and can be seamlessly applied to a wide range of PTMs without requiring architectural modifications. 
Compared with existing PTMs, we achieve up to 6.2× acceleration in inference speed while maintaining classification effectiveness.
Specifically, NetClus has the following three main contributions:
\begin{itemize}
    \item We propose NetClus, a novel framework that integrates clustering-guided fine-tuning and knowledge distillation into existing pre-trained traffic models (PTMs), enhancing both efficiency and adaptability without requiring architectural changes.
    \item NetClus jointly optimizes a clustering-aware feature space and a distilled lightweight encoder to enable high-purity clustering and low-latency classification. A hybrid inference strategy combines fast-path cluster-based prediction with fallback to the original PTM, supporting both efficient inference and discovery of new categories.
    \item Experimental results show that NetClus significantly improves inference speed and classification robustness, while maintaining high effectiveness across diverse encrypted traffic datasets and exhibiting adaptability to previously unseen traffic patterns.
    



\end{itemize}

\section{Related Work}
\subsection{Encrypted Traffic Classification}
We classify existing works on encrypted traffic classification into four categories, based on statistical features, machine learning, deep learning, and pre-training, respectively.

Statistical approaches rely on port-based identification and deep packet inspection (DPI) \cite{moore2005toward}, but these methods fail against modern encryption protocols and dynamic port allocation \cite{qi2009packet}. Subsequent statistical methods utilize handcrafted features like packet size distributions and flow durations, yet struggle with feature engineering overhead and concept drift in evolving network environments \cite{pacheco2018towards}.

Machine learning techniques improve robustness by employing algorithms such as Markov chains for behavioral fingerprinting \cite{flowprint}. Second-order Markov chain models demonstrate effectiveness in identifying encrypted application attributes by analyzing sequential dependencies in traffic flows \cite{korczynski2014markov}. However, these methods remain limited by their dependency on manually designed features and scalability constraints.

Deep learning methods enable automatic feature extraction from raw traffic data. FS-Net \cite{FS-Net} pioneered hierarchical flow sequence modeling using recurrent architectures to capture temporal dependencies in encrypted traffic.
TSRCNN \cite{tscrnn} combined both RNN and CNN to capture spatial and temporal features in and between packets.
However, these methods remained limited by their dependency on manually designed features and scalability constraints.

Pretrained models represent the current frontier, with ET-BERT \cite{etbert} adapting BERT architectures through Same-origin Burst Prediction tasks to model inter-packet relationships. YaTC \cite{yatc} further advanced this paradigm by integrating vision-inspired transformers with multi-level flow representations and masked auto encoding. TrafficFormer \cite{trafficformer}focuses on learning traffic representations through masked language modeling and same-origin data flow tasks and applies data augmentation during fine-tuning to enhance performance in downstream classification tasks. While these pre-trained models reduce dependency on labeled data, they introduce efficiency bottlenecks and lack adaptability to novel traffic patterns.

\begin{figure*}[htbp]  
\centering
\includegraphics[width=\textwidth]{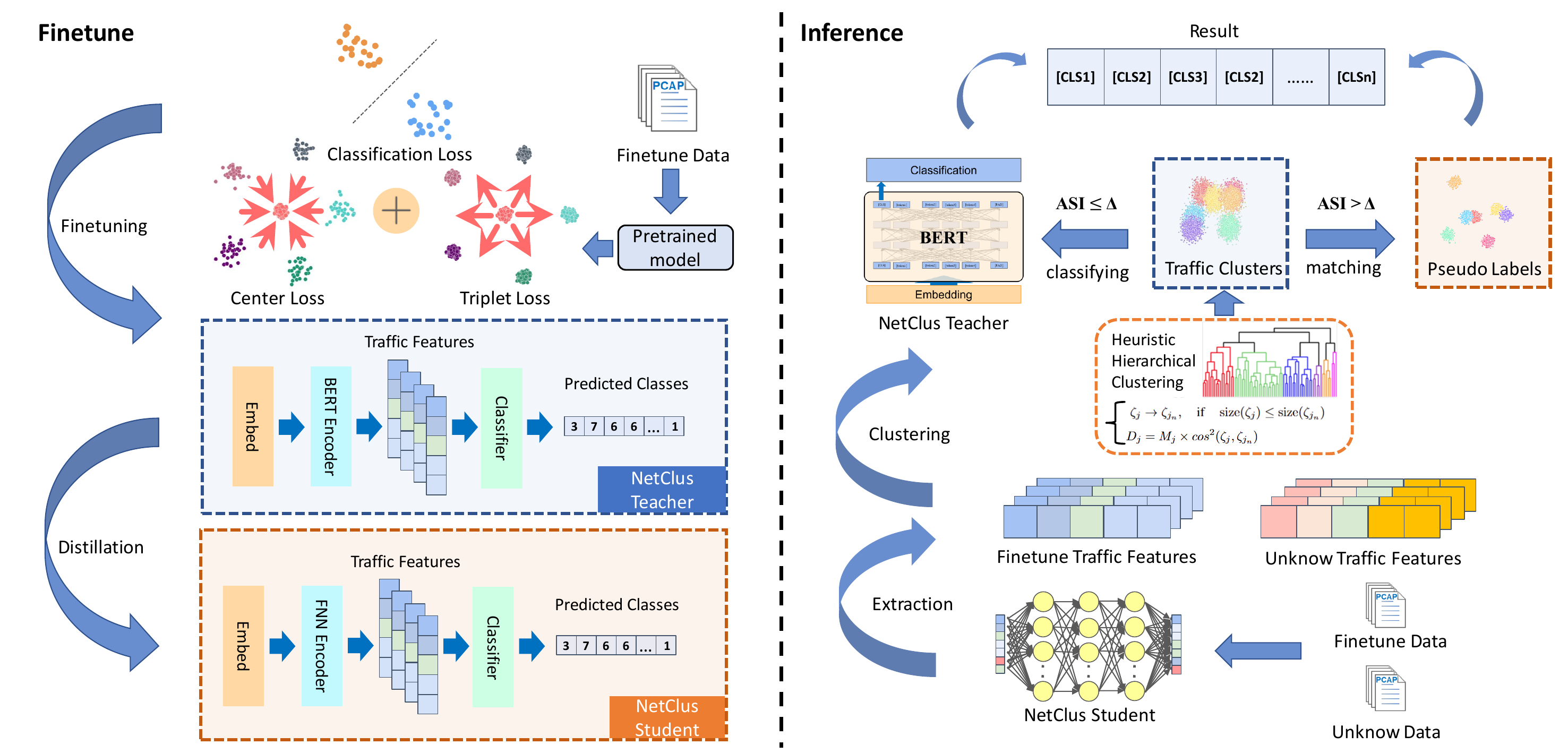}
\caption{The framework of NetClus.}
\label{fig_main}
\end{figure*}

\subsection{Distillation-Based Inference Acceleration}

The deployment of pre-trained models (PTMs) in real-time network traffic classification faces significant challenges due to their slow inference speeds and high computational demands. To address this, knowledge distillation \cite{hinton2015distilling} in the teacher-student framework has emerged as a key technique for compressing large models. Early approaches like {Distilled BiLSTM} \cite{dist_BiLSTM} distill BERT into bidirectional LSTM networks, achieving moderate speedups but with notable accuracy trade-offs. Subsequent innovations include DistilBERT \cite{distilbert}, which reduces BERT's layers for faster inference, and TinyBERT \cite{tinybert}, a 4-layer model leveraging two-stage distillation to balance efficiency and performance. Further advancements culminated in ETED-FNN \cite{ETED}, which distills BERT into a 5-layer fully connected network via Embedding Transformation and Embedding Distillation. This method achieves dramatic acceleration with acceptable performance decline.

However, these general-purpose distillation techniques remain ill-suited for network traffic classification, as they fail to address domain-specific challenges like encrypted payloads, scarce labeled data, and evolving traffic types. To bridge this gap, we propose NetClus, a novel framework optimized for efficient and accurate traffic analysis.

\section{NetClus}
In this section, we present the framework of NetClus and detail its two key stages, as illustrated in Figure \ref{fig_main}. During finetuing, NetClus introduces \textit{clustering-aware knowledge distillation} 
to enable high-purity clustering and low-latency semantic extraction.
During inference, NetClus adopts \textit{hybrid inference accleration} 
to support both efficient label prediction for high-purity clusters and fallback processing for ambiguous through the original pre-trained models. 
Note that NetClus is compatible with arbitrary pre-trained traffic models as input, without requiring structural modification.


\subsection{Clustering-Aware Knowledge Distillation}

To address domain-specific challenges in network traffic analysis, knowledge distillation serves as the foundational step. In the NetClus framework, we first fine-tune the pre-trained model on downstream tasks using a clustering-friendly encoder loss, which optimizes latent space feature vectors for improved cluster separability. Subsequently, we employ a Fully Connected Neural Network (FNN) to distill knowledge from the transformer encoder architecture of pretrained models. This approach achieves significant acceleration in feature vector computation while maintaining clustering purity with minimal degradation.

\subsubsection{Clustering Friendly Encoder Loss.}
Given an input traffic $flow_i$, whose vector representation is $\mathbf{s}_i = \{t_1, t_2, \cdots, t_n\}$ with $n$ tokens, the encoder first embeds it by the pre-trained token embedding layer to an input vector.
Then a transformer architecture processes this input with the attention mechanism, generating the encoded representation $\mathbf{h}_i$ for $flow_i$. The whole process can be represented as follows:
\begin{equation}
\mathbf{h}_i = \mathrm{Transformer}\left( \mathrm{Embed}\left( \mathbf{s}_i\right) \right).
\end{equation}
Based on this representation, we design a clustering-friendly encoder from both classification and clustering perspectives.

1) \textit{Classification Loss.} After getting this representation, the output layer for the downstream task can be defined as:
\begin{equation}
    p_{i, c} = \sigma(\mathbf{h}_i),
\end{equation}
where $\sigma(\cdot)$ is a softmax function to predict the probabilities of the traffic $flow_i$ belonging to traffic category $c$.

Let $y_{i, c}$ denotes the one-hot vector of the ground-truth category of $flow_i$, we can get the origin loss function of the common fine-tuning stage:
\begin{equation}
    \mathcal{L}_{\text{cls}} = -\frac{1}{N} \sum_{i=1}^{N} \sum_{c=1}^{C} y_{i, c} \log(p_{i, c}),
\end{equation}
where $N$ denotes the total number of traffics, $C$ denotes the class number.

2) \textit{Cluster Loss.} Although the current loss function can align the final output layer's predictions with ground-truth labels, this results in representation vectors generated by the encoder lacking optimal clustering properties, thereby hindering subsequent high-purity clustering.
To address this limitation, we introduce a novel loss function for $\mathbf{h}_i$ that enforces two essential clustering characteristics:
\begin{itemize}
    \item \textbf{Intra-cluster compactness}: Minimizing the variance within each cluster.
    \item \textbf{Inter-cluster separation}: Maximizing the distance between different clusters.
\end{itemize}

Specifically, we integrate center loss \cite{center_loss} (minimizing intra-cluster variance) and triplet loss \cite{triplet} (maximizing inter-cluster distance) into the optimization framework.

For \textit{center loss}, we maintain a dynamic class centroid $\mathbf{c}_y$ for each cluster $y$ and minimize the cosine distance between each sample and its corresponding centroid:
\begin{equation}
\mathcal{L}_{\text{center}} = \frac{1}{N} \sum_{i=1}^{N} \left(1 - \cos(\mathbf{h}_i, \mathbf{c}_{y_i})\right),
\end{equation}
where $\mathbf{h}_i$ is the feature representation of the $i$-th sample, $\mathbf{c}_{y_i}$ denotes the centroid of the cluster formed by all data instances with ground truth label $y_i$, and $\cos(\cdot)$ uses cosine function for calculating high-dimensional distance.

To compute the class centroid $\mathbf{c}_{y_i}$, we perform batch-wise updates according to the following procedure: for each mini-batch, we aggregate all feature vectors corresponding to samples with ground-truth class label $y_i$ and compute their arithmetic mean. This vector then directs the historical centroid through a displacement operation. The rate of displacement is determined by the momentum parameter $\alpha \in (0, 1)$, which controls the relative weighting between historical position preservation and current batch alignment. This update process is formally described by the exponential moving average equation:
\begin{equation}
\mathbf{c}_{y_i}^{(t+1)} = \alpha \mathbf{c}_{y_i}^{(t)} + (1 - \alpha) \cdot \frac{\sum_{j:y_j=y_i} \mathbf{h}_i}{|\{j:y_j=y_i\}|}
\end{equation}

For \textit{triplet loss}, we construct triplets of anchor traffic sample $s_a$, positive sample $s_p$, and negative sample $s_n$ to enforce cluster separation by a margin $m$:
\begin{equation}
\mathcal{L}_{\text{triplet}} = \frac{1}{|\mathcal{T}|} \sum_{(a, p, n) \in \mathcal{T}} \max \left( \cos(\mathbf{h}_a, \mathbf{h}_n) - \cos(\mathbf{h}_a, \mathbf{h}_p), m \right),
\end{equation}
where $\mathcal{T}$ denotes the set of valid triplets in the batch, $\mathbf{h}_a$ is the anchor sample, while $\mathbf{h}_p$ and $\mathbf{h}_n$ represent the positive sample from the same cluster and the negative sample from a different cluster, respectively.

Combining center loss and triplet loss, we get cluster loss:
\begin{equation}
    \mathcal{L}_{\text{clus}} = \mathcal{L}_{\text{center}} + \beta \cdot \mathcal{L}_{\text{triplet}},
\end{equation}
where $\beta$ controls the relative contributions of intra-cluster compactness and inter-cluster separation.

3) \textit{Overall Loss.}
In summary, we get a clustering-friendly encoder (CFE) loss by adding classification loss and cluster loss together:
\begin{equation}
    \mathcal{L}_{\text{CFE}} = \mathcal{L}_{\text{cls}} + \lambda \cdot \mathcal{L}_{\text{clus}},
\end{equation}
where $\lambda$ controls the relative contributions of classification and cluster loss.

By employing a clustering-friendly loss function, the traffic features extracted by our fine-tuned model demonstrate improved cluster purity, while classification performance on most datasets also surpasses that of the original model (further experiments can be found in Table \ref{tab:results}).


\subsubsection{FNN Semantic Extractor.}

To expedite feature vector extraction for network traffic, we distill knowledge from the fine-tuned model into a lightweight five-layer Fully Connected Neural Network (FNN). As illustrated in Figure \ref{fig_main}, the frozen fine-tuned model acts as the teacher, and its parameters remain fixed throughout this process. Meanwhile, the FNN model serves as a student model to align with the teacher. Thus, we jointly optimize both the intermediate representations of the encoder layer and the logits output of the output layer. 
Crucially, the clustering-friendly loss introduced during fine-tuning is retained to ensure that the FNN-derived embeddings preserve cluster separability. The overall distillation loss function is formulated as follows:
\begin{equation}
    \hat{L} = \frac{1}{2N} \sum_{i=1}^{N} \left\{ 
      \frac{1}{d} \sum_{j=1}^{d} (\mathbf{\hat{h}}_i^j - \mathbf{h}_i^j)^2 
      + \frac{1}{u} \sum_{r=1}^{u}\sum_{c=1}^{C} p_{i, c}^r \log \frac{p_{i, c}^r}{\hat{p}_{i, c}^r} 
    \right\}
\end{equation}
\begin{equation}
    \left\{
    \begin{aligned}
        \mathbf{\hat{h}}_i &= \mathrm{FNN}\left( \mathrm{Embed}\left( \mathbf{s}_i  \right) \right), \\
        \hat{p}_{i, c} &= \sigma(\hat{\mathbf{h}}_i), 
    \end{aligned}
    \right.
\end{equation}
where $\mathbf{\hat{h}}_i^j$ is the $j$-the dimension of encoded representation of $flow_i$ extracted from FNN, and $\hat{p}_{i, c}^r$ is the predicted probabilities from FNN model of the $r$-th dimension.

The distilled FNN encoder acquires semantically rich, clustering-friendly features while significantly reducing the time required for traffic feature extraction by learning both classification results, the teacher model's encoder outputs, and inter-cluster relations.

\subsection{Hybrid Inference Acceleration}
To enable fast traffic classification based on the distilled FNN features, we combine heuristic hierarchical clustering with Affiliation Strength Index (ASI) validation. Specially, to ensure the high-purity of clusters, we forms a significantly larger number of clusters to capture fine-grained variations in traffic patterns. 
In this way, we not only improve inference effectiveness and efficiency but also support the discovery of novel traffic types.
\begin{itemize}
    \item \textit{Heuristic Hierarchical Clustering}: Generate fine-grained, high-purity clusters through optimized feature grouping, serving as pseudo-labels.
    \item \textit{Affiliation Strength Index Validation}:  For target traffic and pseudo-labeled clusters, we compute their Affiliation Strength Index (ASI) to validate cluster purity. Those high-ASI samples are directly classified via nearest cluster assignment, while low-ASI samples are routed to the pre-trained model for specialized reclassification.
\end{itemize}

\subsubsection{Heuristic Hierarchical Clustering.}
Following the rapid acquisition of traffic feature vectors, we generate abundant pseudo-labels through cluster formation. Given that no predefined cluster count is required and intra-cluster purity is prioritized, we optimize hierarchical clustering via heuristic merging as follows.

During hierarchical clustering, each sample is initially assigned to its own singleton cluster, corresponding to a cluster size of one. To implement heuristic merging, the following connection rule is applied between clusters:
\begin{equation}
\zeta_j \to \zeta_{j_n}, \quad \text{if} \quad \operatorname{size}(\zeta_j) \leq \operatorname{size}(\zeta_{j_n})
\end{equation}
where $\zeta_j$ denotes the $j$-th cluster, $\zeta_{j_n}$ represents its nearest neighboring cluster, and $\operatorname{size}(\cdot)$ is the sample number of the corresponding cluster. This criterion ensures that only smaller clusters are merged into their closest larger counterparts. Through iterative application of this merging process, all clusters eventually consolidate into the root node of the hierarchical tree structure.

To maximize cluster purity, merging large clusters should be minimized as minority-class instances within oversize clusters become indistinguishable, degrading pseudo-label reliability. Consequently, clustering algorithms should prioritize generating uniformly sized clusters to enhance pseudo-label purity \cite{ECHC}.
Therefore, when merging two clusters, we compute the product of their sizes, denoted as $M_j =\operatorname{size}(\zeta_j) \times \operatorname{size}(\zeta_{j_n}) $, and incorporate it into the final distance metric function $D_j$ for cluster merging that:
\begin{equation}
    D_j = M_j \times \cos^2(\zeta_j, \zeta_{j_n}).
\end{equation}

Unlike conventional clustering algorithms with $O(n^2)$ complexity, our method significantly reduces the computation costs. Since the computational process only merges small clusters with their nearest large cluster, each parent node in the hierarchical tree contains at least twice as many samples as its child nodes. This results in a tree height of $O(log n)$, and when distance computation is treated as a constant-time operation, the overall clustering process achieves a time complexity of $O(n log n)$. This yields a near-linear runtime in practice, enabling deployment in high-throughput network settings.

\subsubsection{Affiliation Strength Index Validation.}

To validate whether samples within an input cluster genuinely belong to its assigned pseudo-label category after heuristic clustering, we define the \textit{Affiliation Strength Index} (ASI) of traffic sample $flow_i$ as below: 
\begin{equation}
\text{ASI}(flow_i) = \left( {ratio}_i, \  {strength}_i \right),
\end{equation}
\begin{equation}
\left\{
\begin{aligned}
{ratio}_i &= \frac{ \left| \{ p_j \in \mathcal{N}_k(flow_i) \mid p_j = p_{\text{nearest}} \} \right| }{k}, \\[1ex]
{strength}_i &= \frac{ |d_{\text{inter}}(i) - d_{\text{intra}}(i)| }{ \max\left\{ d_{\text{inter}}(i), d_{\text{intra}}(i) \right\} }.
\end{aligned}
\right.
\end{equation}

Here, $ratio_i$ quantifies the proportion of matching pseudo-labels among the $k$ nearest neighbors surrounding nearest cluster of the sample $flow_i$, while $strength_i$ integrates intra-cluster cohesion and inter-cluster separation via $d_{intra}$ (the average distance from $flow_i$ to other samples in its cluster) and $d_{inter}$ (the average distance from $flow_i$ to samples in the nearest neighboring cluster).

Under the situation that pseudo-label cluster count exceeds the true class count, a high $ratio$ indicates proximity to a true-class cluster centroid, and a high $strength$ implies strong affiliation with the surrounding clusters. When $ratio$ and $strength$ approach 1, $flow_i$ likely resides near a true-class cluster centroid.

Given a discriminant threshold $\Delta = (\gamma, \eta)$, where $\gamma$ and $\eta$ are both hyperparameters, the pseudo-label assignment for sample $\mathbf{s}_i$ is preserved if and only if both conditions are satisfied:
(i) the ratio index exceeds $\gamma$ ($\mathit{ratio}_i > \gamma$), and
(ii) the strength index surpasses $\eta$ ($\mathit{strength}_i > \eta$).
Samples failing to meet this compound criterion are designated for reclassification.
\begin{equation}
\begin{cases} 
ASI(flow_i) > \Delta   &\Rightarrow \text{Retain seudo-label }, \\
ASI(flow_i) \leq \Delta &\Rightarrow \text{Reclassify}.
\end{cases}
\end{equation}
which helps strike a balance between efficiency and adaptability. With high ASI, the predicted label is directly assigned using the cluster's pseudo-label. Otherwise, the sample fallback to the model fine-tuned by $\mathcal{L}_{CFE}$. In summary, for each network traffic, the heuristic hierarchical clustering module determines whether it belongs to a high-purity via the validation by ASI, and filters only small amounts of samples back to the original models, achieving both fast inference and accurate classification.

\begin{table}[t!]
\centering
\small 
\renewcommand{\arraystretch}{0.9} 
\begin{tabular}{@{}>{\raggedright\arraybackslash}p{3cm}p{2cm}rr@{}}
\toprule
\textbf{Datasets} & \textbf{Task} & \textbf{\makecell[c]{Flow\\ Number}} & \textbf{\makecell[c]{Class\\ Number}} \\
\midrule
CSTNET-TLS 1.3 & \makecell[l]{Website\\ Fingerprinting} & 46,372 & 120 \\
ISCX-VPN (Service) & \makecell[l]{Service Type\\ Identification} & 2,242 & 12 \\
ISCX-VPN (App) & \makecell[l]{Application\\ Fingerprinting} & 2,256 & 17 \\
USTC-TFC & \makecell[l]{Malware\\ Detection} & 2,994 & 10 \\
\bottomrule
\end{tabular}
\caption{Network traffic datasets for experimental evaluation}
\label{tab:datasets}
\end{table}

\begin{table*}[t!]
\centering
\small
\begin{threeparttable}
\begin{tabular}{@{}l *{4}{c c c} c @{}}
\toprule
\multirow{2}{*}{\textbf{Methods}} & 
\multicolumn{3}{c}{\textbf{CSTNET-TLS 1.3}} & 
\multicolumn{3}{c}{\textbf{ISCX-VPN (Service)}} & 
\multicolumn{3}{c}{\textbf{ISCX-VPN (App)}} & 
\multicolumn{3}{c}{\textbf{USTC-TFC}} & 
\multirow{2}{*}{\textbf{Avg. F1}} \\
\cmidrule(lr){2-4} \cmidrule(lr){5-7} \cmidrule(lr){8-10} \cmidrule(lr){11-13}
& \textbf{PR} & \textbf{RC} & \textbf{F1} 
& \textbf{PR} & \textbf{RC} & \textbf{F1} 
& \textbf{PR} & \textbf{RC} & \textbf{F1} 
& \textbf{PR} & \textbf{RC} & \textbf{F1} & \\
\midrule
FlowPrint & 11.90 & 11.79 & 10.27 & 63.86 & 63.46 & 62.27 & 58.87 & 56.82 & 54.51 & 76.09 & 67.48 & 68.63 & 48.92 \\
\midrule
DeepPacket & 40.95 & 33.63 & 32.89 & 78.61 & 75.38 & 76.50 & 60.47 & 48.53 & 51.75 & 97.09 & 97.18 & 97.13 & 64.57 \\
LSTM-Att & 26.55 & 19.26 & 18.67 & 68.97 & 63.11 & 65.21 & 60.11 & 37.93 & 42.36 & 95.14 & 94.88 & 95.00 & 55.31 \\
DataNet & 0.01 & 0.83 & 0.02 & 57.17 & 34.57 & 40.36 & 23.13 & 13.17 & 13.84 & 88.34 & 82.08 & 83.96 & 34.55 \\
\midrule
YaTC & 86.61 & 84.98 & 84.59 & 87.77 & 86.82 & 86.48 & 68.87 & 47.85 & 48.99 & \textbf{97.47} & \underline{97.44} & \underline{97.43} & 79.37 \\
ET-BERT & 84.85 & 83.91 & 84.12 & 90.99 & 90.81 & 90.40 & 70.46 & \underline{67.31} & 68.40 & \underline{97.39} & \textbf{97.68} & \textbf{97.46} & 85.10 \\
TrafficFormer & \textbf{88.40} & \textbf{88.29} & \textbf{88.14} & \textbf{93.58} & \textbf{92.23} & \textbf{92.76} & \underline{71.58} & \textbf{69.02} & \textbf{69.94} & 97.11 & 97.33 & 97.17 & \textbf{87.00} \\
\midrule
NetClus & \underline{87.44} & \underline{86.67} & \underline{86.83} & \underline{92.58} & \underline{92.17} & \underline{92.23} & \textbf{72.02} & 66.68 & \underline{68.73} & 97.29 & 97.37 & 97.27 & \underline{86.27} \\
NetClus\_large \textsuperscript{1} & 88.45 & 88.31 & 88.18 & 92.65 & 92.42 & 92.44 & 72.39 & 67.70 & 69.57 & 97.31 & 97.60 & 97.35 & 86.89 \\
\bottomrule

\multicolumn{14}{l}{\textsuperscript{1} NetClus\_large is a large-parameter classification model obtained by fine-tuning TrafficFormer \cite{trafficformer} using CFE loss.}

\end{tabular}
\end{threeparttable}
\caption{Performance comparison (\%) on network traffic classification benchmarks with average F1 score.}
\label{tab:results}
\end{table*}

\subsection{Dynamic Adaptation to New Traffic Types}
Beyond effective and efficient inference, NetClus also supports dynamic adaptation to previously unseen traffic categories, which is a critical capability for open-world network environments. This can be easily achieved by leveraging existing clustering-based signals without incurring additional computational overhead.

While determining whether a sample belongs to its corresponding pseudo-label using ASI, we can simultaneously employ this criterion to detect potential novel categories. Specifically, for a given cluster $clus_k$, we define  $ASI(clus_k) = (ratio_k, strength_k)$, where $ratio_k$ is computed based on the cluster centroid, and $strength_k$ represents the average strength of samples within the cluster. A cluster exhibiting a low ratio but high strength is noteworthy, i.e., they exhibit strong internal similarity yet fail to align with known categories. Such clusters have high probabilities and can be identified as novel network traffic types. 

This approach enables lightweight novelty detection in fast inference, allowing NetClus to recognize emerging traffic patterns without incurring the cost of additional labeling, retraining, or external detectors. It complements the fallback process by distinguishing between ambiguous know traffic and truly novel traffic, thereby supporting traffic classification in dynamic network environments.


\section{Experimental Results}

\subsection{Settings}
\subsubsection{Datasets} 
We conduct comprehensive evaluations on four benchmark network traffic datasets: ISCX-VPN (Service) \cite{ISCX}, ISCX-VPN (App) \cite{ISCX}, CSTNET-TLS 1.3 \cite{etbert}, and USTC-TFC \cite{USTC}.
Table \ref{tab:datasets} shows the information on network traffic datasets for experimental evaluation.
These datasets encompass four specific finetuning tasks: application fingerprinting, service type identification, website fingerprinting, and malware detection.
\textbf{CSTNET-TLS 1.3} includes traffic data aggregated from 120 websites utilizing the TLS 1.3 protocol. \textbf{ISCX-VPN (Service)} and \textbf{ISCX-VPN (App)} datasets encapsulate network traffic generated by diverse application behaviors over Virtual Private Networks (VPNs), enabling classification into specific services and applications. The \textbf{USTC-TFC} dataset comprises traffic from 10 benign software applications and 10 malware samples.

\subsubsection{Baselines}
We select seven baselines, including one machine learning based method (FlowPrint \cite{flowprint}), three deep learning based methods (DeepPacket \cite{deeppacket}, LSTM-Att \cite{LSTM_Att}, Datanet \cite{Datanet}), and three pre-trained model based methods (ET-BERT \cite{etbert}, YaTC \cite{yatc}, and TrafficFormer \cite{trafficformer}).

\subsubsection{Implementation Details}  

The experimental configuration employs an 8:1:1 ratio for partitioning datasets into training, validation, and testing subsets, ensuring model evaluation while preventing overfitting. During finetuning, we utilize 10 epochs with a batch size of 256, implementing class-balanced sampling by capping each category up to 5,000 flows to mitigate class imbalance biases. We use the first 5 packets in a flow with 128-byte payloads. Specifically, for knowledge distillation, the process spans 20 epochs since the student model is trained from scratch. All experiments are deployed on NVIDIA H800 GPUs using PyTorch 2.7.0.

\begin{figure}[t]
\centering
\begin{subfigure}{0.45\linewidth}
    \centering
    \includegraphics[width=\linewidth]{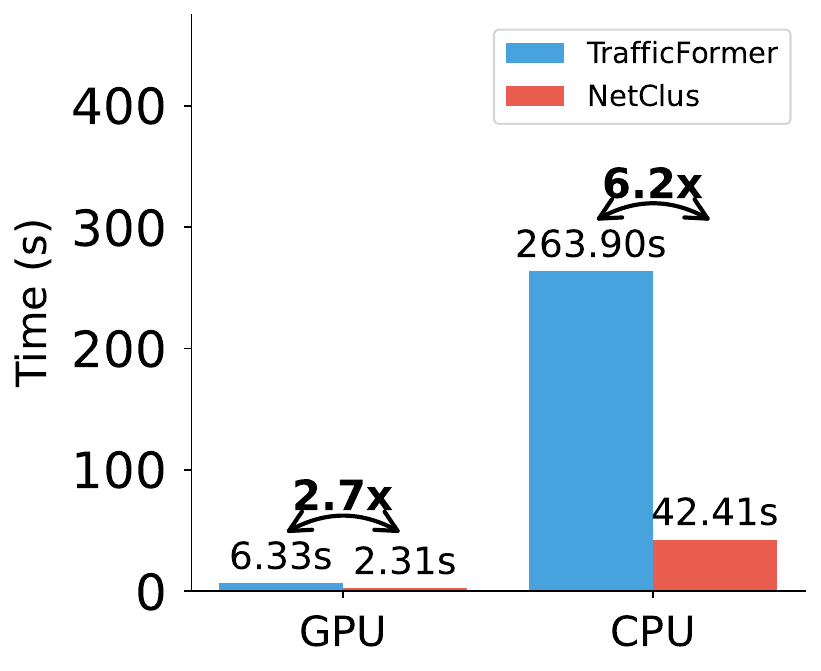}
    \caption{Trafficformer vs NetClus}
    \label{fig:tf_netclus}
\end{subfigure}
\hfill
\begin{subfigure}{0.45\linewidth}
    \centering
    \includegraphics[width=\linewidth]{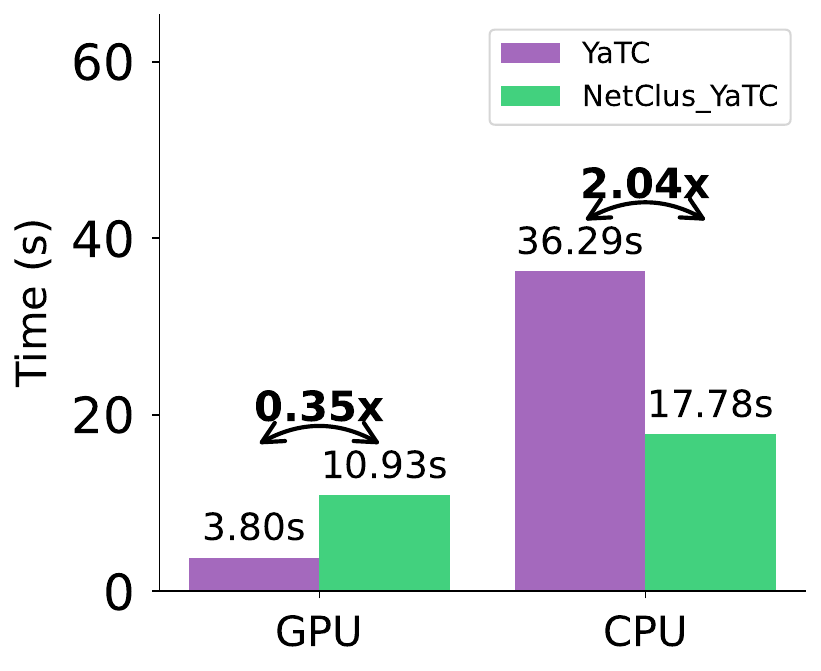}
    \caption{YaTC vs NetClus\_YaTC}
    \label{fig:ytc_netclus}
\end{subfigure}
\caption{Inference time comparison of NetClus across diverse pre-trained models and hardware}
\label{fig:time_comparison}
\end{figure}

\subsubsection{Evaluation Metrics}  

Let $T_p$, $T_n$, $F_p$, and $F_n$ represent True Positive, True Negative, False Positive, and False Negative. The performance is quantified using:  
\begin{align}
  \text{Precision} &= \frac{T_p}{T_p + F_p}, \qquad
  \text{Recall}    = \frac{T_p}{T_p + F_n} \label{eq:both} \\
  F_1 &= 2 \times \frac{\text{Precision} \times \text{Recall}}{\text{Precision} + \text{Recall}} \label{eq:f1}
\end{align}  

During calculation, macro-averaging is applied for multi-class scenarios to ensure equal class weighting.  

\subsection{Overall Performance Comparisons}

As shown in Table \ref{tab:results}, our NetClus method achieves at least second-best performance across all evaluated datasets except USTC-TFC. Furthermore, the TrafficFormer\_cluster model, fine-tuned with CFE loss, outperforms the original TrafficFormer on all datasets except ISCX-VPN(Service). The performance gap between NetClus and baseline models remains within approximately 1\%. This demonstrates that our NetClus method, through knowledge distillation and clustering comparison techniques, achieves efficacy comparable to the original pre-training models.

Moreover, pre-training models consistently outperform deep learning approaches, which in turn surpass traditional machine learning methods. This advantage is particularly notable on the CSTNET-TLS 1.3 dataset, which employs advanced payload encryption, and pre-training models demonstrate substantial improvements over those methods.

\subsection{Inference Efficiency Comparisons}
To evaluate the inference acceleration efficacy of NetClus, we conduct experiments on the USTC-TFC dataset, comparing inference speed and performance differences between NetClus and baseline methods (i.e., TrafficFormer and YaTC).
As shown in Figure \ref{fig:time_comparison}, when optimizing large-parameter pre-trained models (i.e., TrafficFormer), NetClus achieves significant acceleration across computational platforms: a 2.7× speedup ratio on GPU and 6.2× on CPU, while slightly outperforming TrafficFormer in classification performance.
This underscores NetClus’s effectiveness in accelerating complex models.

For compact models like YaTC, NetClus incurs overhead due to clustering parameter calculations, reducing speed to one-third of the original method for small-scale data. However, under CPU computation, NetClus exhibits only a 1.7× time increase, whereas YaTC suffers a ~10× slowdown. Thus, while NetClus demonstrates limited gains for small models trained on minimal data, its processing speed becomes multiplicatively superior as traffic volume scales.

\begin{figure}[t!] 
\centering
\includegraphics[width=0.95\linewidth,  clip]{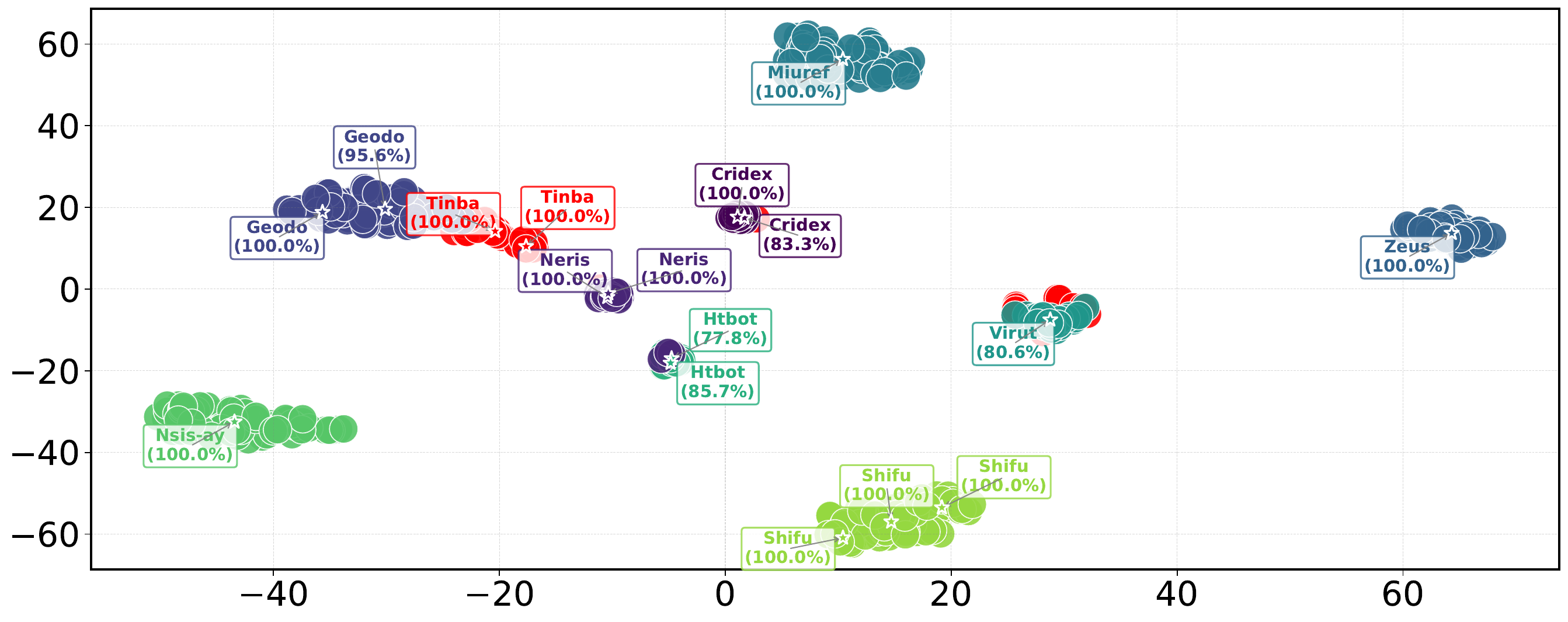}
\caption{Visualization of NetClus clustering results. Novel traffic categories (Tinba) are colored in red to emphasize their distinction from known malware families.}
\label{fig:new type}
\end{figure}

\begin{figure}[t!] 
\centering
\includegraphics[width=0.95 \linewidth,  clip]{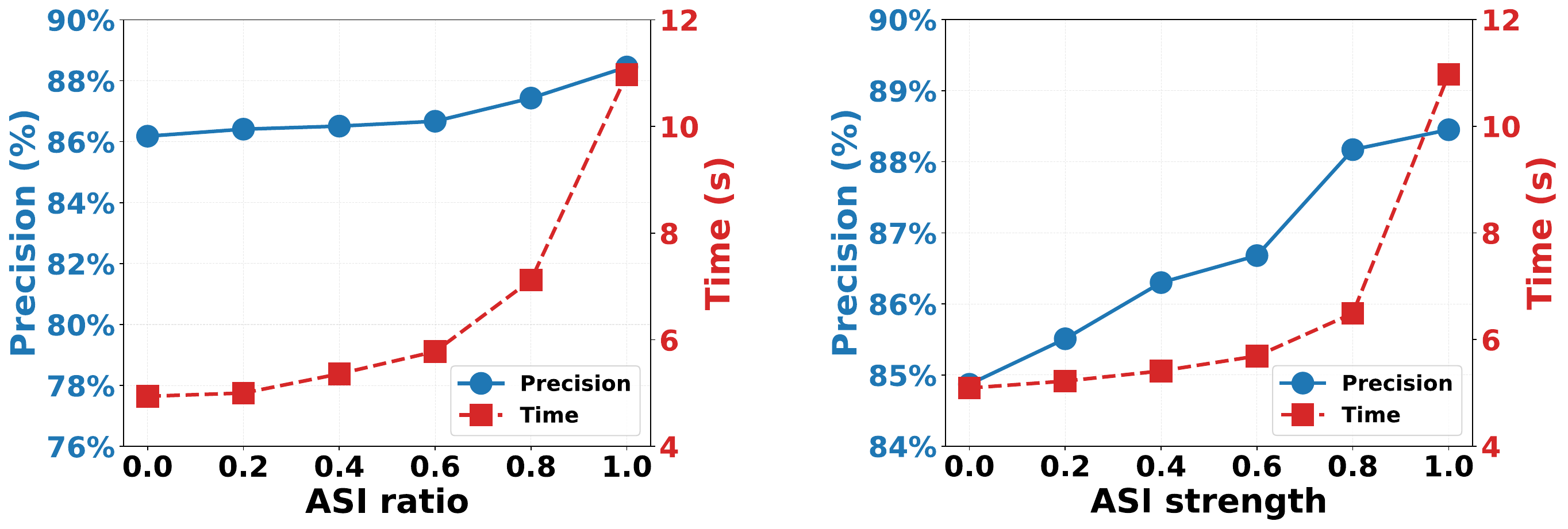}
\caption{ASI ratio (left) and strength (right) analysis.}
\label{fig:delta}
\end{figure}

\subsection{New Type Identification}

During traffic classification, NetClus inherently clusters input network traffic. This enables direct identification of novel traffic types using the ASI coefficient of clusters without additional operations. A lower ASI value indicates a higher probability of a new traffic type, while a higher value suggests membership in a known type.

We validate this on the USTC-TFC dataset using only 9 predefined traffic classes during training, explicitly excluding Tinba malware traffic from the training set. Then we introduce the unseen class during testing. Results in Figure \ref{fig:new type} demonstrate that NetClus effectively distinguishes novel traffic types through cluster-based ASI analysis.

\subsection{ASI Discriminant Study}
As a critical component of the NetClus framework, the ASI discriminant factor serves as a hyperparameter influencing both classification efficacy and inference efficiency. We investigate its impact by fixing ratio=0.5 and strength=0.5 while varying the other parameter. As illustrated in Figure \ref{fig:delta}, both classification accuracy and time cost increase with elevated ratio and strength values. Precision exhibits a nearly linear progression with parameter escalation, while time cost demonstrates exponential growth. These experimental results confirm that the ASI index effectively distinguishes between easily classifiable and challenging traffic clusters, thereby optimizing inference acceleration.

\begin{table}[!t]
\centering
\footnotesize
\begin{tabular}{@{}l c c c >{\centering\arraybackslash}p{1.8cm}@{}}
\toprule
\multirow{2}{*}{\textbf{Method}} & \multicolumn{3}{c}{\textbf{CSTNET-TLS 1.3}} & \multirow{2}{*}{\textbf{Time (s)}} \\
\cmidrule(lr){2-4}
 & \textbf{PR} & \textbf{RC} & \textbf{F1} & \\
\midrule
NetClus  & 87.44 & 86.67 & 86.83 & 5.93 \\
NetClus w/o Clus & 79.36 & 77.80 & 78.01 & 2.13 \\
NetClus\_large w/ CFE & 88.50 & 88.36 & 88.23 & 14.90 \\
Trafficformer & 88.25 & 88.20 & 88.01 & 17.32 \\
\bottomrule
\end{tabular}
\caption{Ablation studies on clustering and distillation.}
\label{tab:combined_ablation}
\end{table}

\subsection{Ablation Study}
In the ablation study, we focus on three core components: CFE loss, model distillation, and hybrid inference acceleration, comparing on precision and time cost.

As shown in Table \ref{tab:combined_ablation}, we check the effectiveness of the clustering method by removing the cluster-matching module and using direct classification, which causes a ~10\% performance drop, underscoring the critical role of cluster-matching in final inference performance. Besides clustering, we employ Trafficformer, NetClus\_large, and NetClus as feature extractors to verify the effectiveness of distillation and CFE loss. Without CFE loss, NetClus\_large exhibits a 0.2\% degradation in classification performance from classifying directly. This decline stems from erroneous classification results introduced during pseudo-label matching. With CFE loss, NetClus\_large achieves a 22\% speed up while outperforming the original model. These improvements stem from the CFE loss, which optimizes traffic features for clustering algorithms, yielding clusters with higher ASI and reducing reprocessing cases. Compared to Trafficformer, NetClus attains a 2.92× inference speedup with less than 1\% precision degradation, demonstrating the efficacy of the distillation process.


\section{Conclusion}

In this paper, we propose NetClus, an efficient inference acceleration framework for pre-trained traffic classification models. The method employs CFE loss to extract clustering-optimized traffic features, rapidly obtains high-purity clusters through model distillation and heuristic clustering, and achieves inference acceleration via cluster-to-pseudolabel matching. NetClus exhibits broad applicability for accelerating diverse pre-trained traffic classification models. Evaluated across four benchmark network traffic datasets, experimental results demonstrate that NetClus achieves up to 6.2× speedup with $< 1\%$ accuracy degradation compared to state-of-the-art methods. Furthermore, NetClus enables recognition of novel traffic types without introducing additional computational overhead.

\bibliography{aaai2026}



\end{document}